\documentclass[twocolumn,nofootinbib]{revtex4}

\usepackage{graphicx}
\usepackage{color}
\usepackage{enumerate}
\usepackage{amssymb}
\usepackage{amsmath}
\usepackage{amsfonts}
\usepackage{bbm}
\usepackage{dsfont}
\def\be{\begin{equation}} 
\def\ee{\end{equation}}

\begin{document}

\title{Barrier penetration with a finite mesh method}

\author{K. Hagino}
\affiliation{ 
Department of Physics, Kyoto University, Kyoto 606-8502,  Japan} 

\begin{abstract}
A standard way to solve a Schr\"odinger equation is to discreteize the radial coordinates 
and apply a numerical method for a differential equation, such as the Runge-Kutta method or 
the Numerov method. Here I employ a discrete basis formalism based on a finite mesh method 
as a simpler alternative, 
with which the numerical computation can be easily implemented by ordinary linear algebra 
operations. 
I compare the numerical convergence of the
Numerov integration method to the finite mesh method for calculating
penetrabilities of a one-dimensional potential barrier, and show 
that the latter approach has better convergence properties.
\end{abstract}

\maketitle

\section{Introduction}

In most of physics problems, a Schr\"odinger equation cannot be solved 
analytically but has to be solved numerically. For a bound state problem, 
one may expand wave functions on some finite basis and diagonalize the resultant 
Hamiltonian matrix. Alternatively, one may discretize the radial coordinates and 
successively obtain a wave function at the mesh points with e.g., the Runge-Kutta 
method or the Numerov method \cite{koonin}. 

A yet different method, referred to as a discrete-basis formalism\footnote{
Even though the term \lq\lq discrete-basis formalism" was not introduced in Ref. \cite{FBA18}, 
the method given in Ref. \cite{FBA18} is equivalent to the discrete-basis formalism 
shown in later publications \cite{BY2019,YBF2020,YBF2021}.},
has been proposed in Ref. \cite{FBA18}. 
In this method, one first forms a Hamiltonian matrix based on discretized radial 
meshes and solve it with a linear algebra with appropriate boundary conditions. 
An advantage of this method is that the method is well compatible with a many-body 
Hamiltonian, in particular in a configuration-interaction formulation \cite{BH2022,BH2023,UH2023}. 
Notice that the discrete-basis formalism is referred to as a 3D mesh method in the context of 
nuclear density functional theory\cite{Davies80,ev8,ev8-2,tanimura2015,ren2017,li2020}. 

Even though the discrete-basis formalism has been applied 
to an induced fission problem \cite{FBA18,BY2019,YBF2020,YBF2021,BH2023}, 
its applicability has not yet been clarified, at least for a scattering problem. 
In this paper, I therefore apply the discrete-basis formalism to a simple 
one-dimensional barrier penetration problem, and carry out a comparative study 
of the numerical accuracy. To this end, I shall consider a Gaussian barrier and 
compare the penetrabilities obtained 
with the discrete-basis formalism to those with the standard Numerov method. 

The paper is organized as follows. In Sec. II, I will detail the discrete-basis formalism 
for a one-dimensional problem. In Sec. III, I will apply it to a barrier penetration of 
a one-dimensional Gaussian barrier and discuss the applicability of the discrete-basis formalism. 
I will then summarize the paper in Sec. IV. 

\section{Discrete-basis formalism for barrier penetration}

Consider a one-dimensional system for a particle with mass $m$ under a potential $V(x)$. 
The Hamiltonian for this system reads,
\begin{equation}
H=-\frac{\hbar^2}{2m}\frac{d^2}{dx^2}+V(x).
\label{eq:H}
\end{equation}
I discretize the radial coordinate as $x_i=x_{\rm min}+(i-1)\Delta x$, 
and consider the model space from $x_1=x_{\rm min}$ to 
$x_N\equiv x_{\rm max}$. 
Using the 3-point formula for the kinetic energy in $H$, 
the Hamiltonian (\ref{eq:H}) is transformed to a matrix form of 
\begin{equation}
H_{ij}=-t\delta_{i,j+1}+(2t+V_i)\delta_{i,j}-t\delta_{i,j-1},
\end{equation}
where $t$ is defined as $t=\frac{\hbar^2}{2m(\Delta x)^2}$ and $V_i\equiv V(x_i)$. 
The wave function $\phi_i\equiv\phi(x_i)$ then obeys 
\begin{equation}
    -t\phi_0\delta_{i,1}+\sum_{j=1}^NH_{ij}\phi_j-t\phi_{N+1}\delta_{i,N}=E\phi_i.
\label{eq:matrix}
\end{equation}

In the absence of the potential $V$, the wave function $\phi_n^{(0)}$ 
obeys the equation
\begin{equation}
-t(\phi_{n+1}^{(0)}-2\phi_{n}^{(0)}+\phi_{n-1}^{(0)})=E\phi_{n}^{(0)}.
\label{eq:free}
\end{equation}
I consider a free-particle solution given by 
\begin{equation}
\phi_n^{(0)}\propto e^{-ikn\Delta x}-e^{ikn\Delta x}.
\end{equation}
Substituting this into Eq. (\ref{eq:free}), one finds 
\begin{equation}
\cos(k\Delta x)=1-\frac{E}{2t}.
\label{eq:k}
\end{equation}

In the presence of the potential $V$, I 
consider the case where the particle is incident from the left hand side of the potential.  
Assuming that the potential $V$ almost vanishes at $x_{\rm max}$, the wave 
function $\phi_{N+1}$ is given by $\phi_{N+1}=e^{ik\Delta x}\phi_N$.  
Substituting this into Eq. (\ref{eq:matrix}), one finds
\begin{equation}
\phi_i=[(\tilde{H}-E)^{-1}]_{i1}t\phi_0\equiv G_{i1}t\phi_0,
\end{equation}
where $\tilde{H}$ is defined as $\tilde{H}_{ij}=H_{ij}-te^{ik\Delta x}\delta_{i,N}\delta_{j,N}$ 
and the Green's function $G$ is defined as $G=(\tilde{H}-E)^{-1}$. 

Assuming that the potential $V(x)$ is negligible at $x=x_1$ and $x_2$, the wave functions 
at these points are given as linear superpositions of $e^{\pm ikn\Delta x}$ 
with $n=1$ and 2, respectively. I parameterize the coefficients of the linear superpositions in terms of 
$t$ and the wave function $\phi_0$ as, 
\begin{eqnarray}
\phi_1&=&(Ae^{ik\Delta x}+Be^{-ik\Delta x})t\phi_0, 
\label{eq:phi1}
\\
\phi_2&=&(Ae^{2ik\Delta x}+Be^{-2ik\Delta x})t\phi_0. 
\label{eq:phi2}
\end{eqnarray}
This is equivalent to assume 
\begin{equation}
G_{11}=Ae^{ik\Delta x}+Be^{-ik\Delta x}. 
\label{eq:G11}
\end{equation}
Substituting Eqs. (\ref{eq:phi1}) and (\ref{eq:phi2}) into Eq. (\ref{eq:matrix}) 
and using Eq. (\ref{eq:k}), one finds 
\begin{equation}
    Ae^{2ik\Delta x}+Be^{-2ik\Delta x}=2\cos(k\Delta x)G_{11}-\frac{1}{t}. 
\end{equation}
Combining this with Eq. (\ref{eq:G11}), one finds 
\begin{eqnarray}
    A&=&\frac{e^{-ik\Delta x}}{e^{ik\Delta x}-e^{-ik\Delta x}}\,(e^{ik\Delta x}G_{11}-1/t), \\
    B&=&-\frac{e^{ik\Delta x}}{e^{ik\Delta x}-e^{-ik\Delta x}}\,(e^{-ik\Delta x}G_{11}-1/t).
\end{eqnarray}
Writing the wave function $\phi_N$ as $\phi_N=G_{N1}t\phi_0\equiv Te^{ik\Delta x}t\phi_0$, 
the penetrability $P(E)$ reads,
\begin{equation}
P(E)=\left|\frac{T}{A}\right|^2=\left|\frac{2\sin(k\Delta x)G_{N1}}{e^{ik\Delta x}G_{11}-1/t}
\right|^2.
\end{equation}

\section{Penetrability of a Gaussian barrier}

\begin{figure}[t]
\includegraphics[width=7cm]{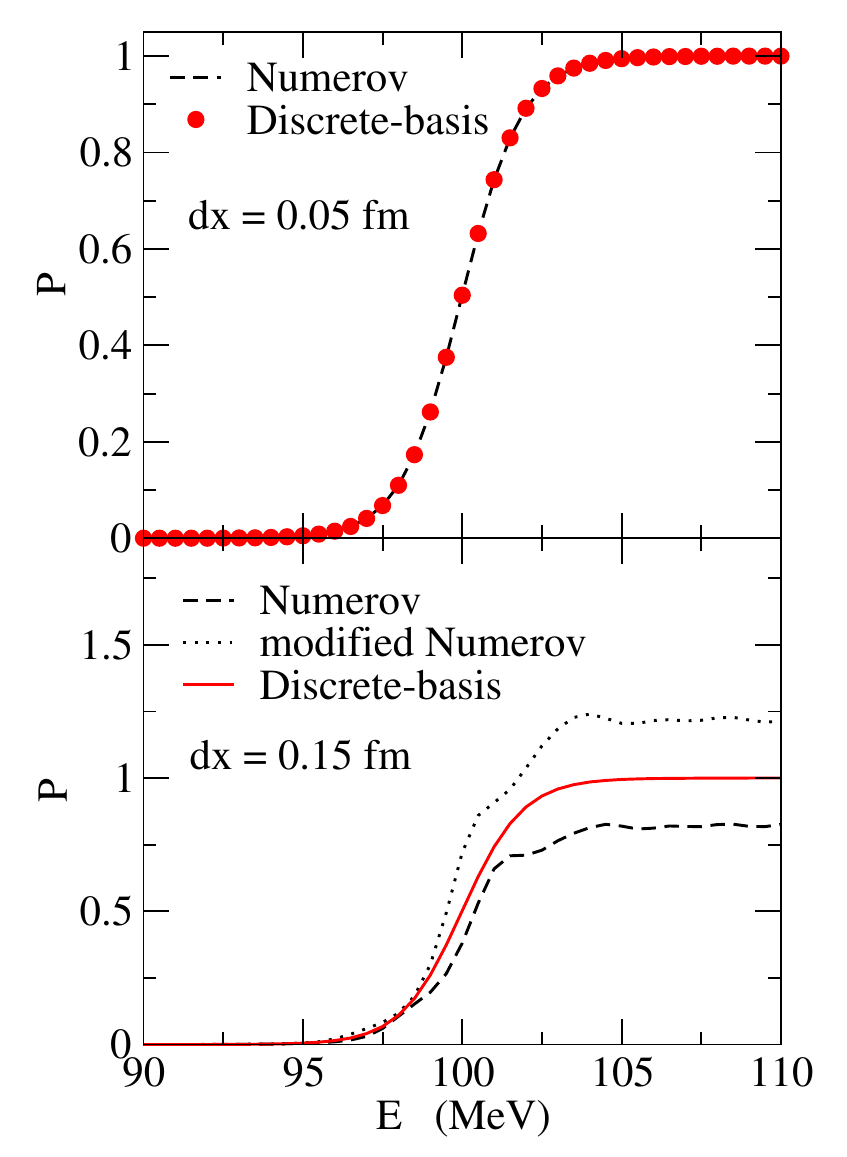}
\caption{
The penetraibilities of a Gaussian barrier given by Eq. (\ref{eq:Gauss}) with $V_0=100$ MeV and $s=$ 2 fm. 
The mass is set to be $m=29m_N$, where $m_N$ is the nucleon mass. 
The upper panel is obtained with the Numerov method (the dashed line) and the discrete-basis formalism (the filled circles) 
with the mesh size of $\Delta x=0.05$ fm. 
On the other hand, the lower panel shows the results of the Numerov method (the dashed line), the
modified Numerov method (the dotted line), and the discrete-basis formalism (the solid line) 
with the mesh size of $\Delta x=0.15$ fm. 
}
\end{figure}

Let us now numerically evaluate the penetrability for a given potential. For this purpose, I  
consider a Gaussian potential, 
\begin{equation}
V(x)=V_0\,e^{-x^2/2s^2}.
\label{eq:Gauss}
\end{equation}
Following Refs. \cite{dasso83,dasso83b,HB04}, 
the parameters are chosen to be $V_0=100$ MeV and $s=2$ fm together with $m=29m_N$, where $m_N$ 
is the nucleon mass, to mimic the fusion reaction of $^{58}$Ni+$^{58}$Ni. 
We set $x_{\rm min}=-10$ fm and $x_{\rm max}=10$ fm. 

The upper panel of Fig. 1 shows 
the penetrabilities of the Gaussian barrier obtained with $\Delta x=0.05$ fm. 
The dashed line and the solid circles denote the results with the standard Numerov method and the 
discrete-basis formalism, respectively. The value of $\Delta x$ is small enough in this case, and 
both the 
method lead to accurate results. 
The lower panel shows the results with a larger value of $\Delta x$, that is,  $\Delta x=0.15$ fm. 
In this case, the numerical error is significantly large with the Numerov method: 
the penetrabilities do not reach unity even at energies well above the barrier (see the dashed line). 
This is the case also with the modified Numverov method \cite{ccfull}, with which the 
penetrability 
even exceeds unity at high energies with a non-monotonic behavior 
(see the dotted line). In marked contrast, the results with the discrete-basis formalism is rather 
robust and the penetrabilities are almost the same as the one with $\Delta x=0.05$ fm shown in the upper panel. 
Notice that the discrete-basis formalism employs the simple 3-point formula for 
the kinetic energy, while a more sophisticated formula is used in the Numerov and the 
modified Numerov methods. 
Yet, it is interesting to notice that the discrete-basis method is numerically more stable than the Numerov and the modified Numerov methods. 
We point out that $\Delta x$ cannot be taken larger than $\sqrt{2\hbar^2/Em}$, though. 
If $\Delta x$ exceeds this value, the right hand side of Eq. (\ref{eq:k}) exceeds unity 
and the wave number $k$ cannot be defined unless it is extended to a complex number. 

\section{Summary}

I examined the applicability of the discrete-basis method for a reaction theory. 
To this end, I considered barrier penetration of a one-dimensional Gaussian barrier. 
It was demonstrated that the discrete-basis method provides a more accurate and stable 
method than the standard Numerov method.  
This property may be helpful in obtaining numerically stable solutions of 
coupled-channels equations \cite{HOM2022,HT2012}. 

The discrete-basis formalism has a good connection to a many-body Hamiltonian. 
As a matter of fact, there have been several applications of this method to microscopic 
descriptions of induced fission. In such applications, absorbing potentials, or imaginary 
energies, are introduced to a model Hamiltonian, and the absorbing probability is computed 
with the so called Datta formula \cite{FBA18}. Even though the model setup is somewhat different 
from a barrier problem in one-dimension, in which there is no absorbing part in the Hamiltonian, 
the conclusion in this paper would remain the same in the fission problem as well. 

\begin{acknowledgments}
The author thanks G.F. Bertsch for helpful discussions and for a careful reading of the manuscript. 
This work was supported in part by
JSPS KAKENHI Grant Numbers JP19K03861 and JP23K03414.
\end{acknowledgments}

\end{document}